\documentstyle[12pt]{article}
\begin{document}
\setlength{\unitlength}{1mm} \pagestyle{plain} 

{\large {\bf 1}} \quad {\large {\bf Introduction}} \bigskip

In the present
paper, we consider an interface between two liquid layers subject to
horizontal harmonic vibrations, of amplitude $a$ and frequency $\omega$.

In experimental works by Bezdeneznykh {\it et al.}~\cite{bezdeneznykh}
and by Wolf~\cite{wolf69} for a long horizontal reservoir filled
with two immiscible viscous fluids, an interesting
phenomenon was found at the interface: the horizontal vibrations lead to
the formation of a steady relief. This formation mechanism
 has a threshold nature; it is
noteworthy that such a wavy relief appears on the interface only if the
densities of the two fluids are close enough, i.e. it does not appear for the
liquid/gas interface (free surface).
The interface is absolutely unstable if the heaviest fluid
occupies the upper layer; i.e., the horizontal vibration does
not prevent the evolution of Rayleigh-Taylor instability, in contrast to the
 vertical one which under certain conditions suppresses its evolution.
A theoretical description of this phenomenon was provided by Lyubimov
\& Cherepanov~\cite{lyubimov} within the framework of a high frequency
(of the vibration) approximation and an averaging procedure; they found that a horizontal
vibration leads to a quasi-equilibrium state i.e., a state where
the mean motion is absent but the interface oscillates with a small
amplitude (of the order of magnitude of the cavity displacement)
with respect to the steady relief. In particular, they found that the
relief
(in the case of a heavy fluid occupying the bottom layer) with
finite wavelength is not possible for any thickness of fluid layers.
In fact, the relief with finite wavelength arises only for considerably
thick layers of height
$h>\left[3\alpha/\left(g\left(\rho_1-\rho_2\right)\right)\right]^{1/2}$,
where $\alpha$ is the coefficient of surface tension, $\rho_1$ and $\rho_2$
are the densities of lower and upper layer, respectively; $g$ is the gravity
acceleration.
In the approach~\cite{lyubimov}, two parameters were assumed to be
asymptotically small simultaneously:
(i) the dimensionless thickness of the viscous skin-layers
$\delta =h^{-1}\sqrt{\nu /\omega}$, $\nu$ being the kinematic
viscosity and
(ii) the dimensionless amplitude of the vibration $\epsilon=a/h$.
But in the limiting case $\epsilon \rightarrow 0$,
the possibility of description of parametric resonant effects is absent and
only the basic instability mode (Kelvin--Helmholtz, of two counter
flows) remains.

In this investigation, the condition $\epsilon \rightarrow 0$ is dropped and
the vibration frequency is considered as relatively low, thus the averaging
method could no longer be applied.

\bigskip
\bigskip
\noindent {\large {\bf 2}} \quad {\large {\bf Problem formulation}}
\bigskip

Let us consider the system of two immiscible, incompressible liquids filling rectangular
cavity of length $L$ and height $h$.
In the state of rest 
the heavy liquid (of density $\rho_1$) occupies the bottom region of height $h_{1}$,
and the light liquid (of density $\rho_2$) -- the upper region of height $h_{2}$
($h=h_1+h_2$).
We choose Cartesian coordinate system in such a way that the $x,y$-axis lie in horizontal plane,
the $z$-axis is directed vertically, $z=0$ corresponds to the unperturbed interface.
Let the cavity perform harmonic oscillation along the $x$-axis,
with the amplitude $a$ and frequency $\omega$, and
let $\vec \gamma$ be the unit
vector along the $z$-axis, while $\vec j$ along the vibration axis (Fig. 1).

The waves are generated on the interface due to such oscillation, and propagate
from the vertical endwalls to the central part of the cavity.
But, if viscosities of the liquids are not too small, such a waves are damped on relatively
short distances from the endwalls.
For example, for transformer oil and glycerine in Earth gravity field and 
angular frequencies of the 
vibration
of the order of magnitude $10^2 Hz$, the damping length $\;l_d$ is of the order of 
magnitude of 1 $mm$.
In the following, we assume that the condition $L \gg l_d$ is satisfied and
we consider only the central part of the cavity, far away from the endwalls,
which the interface waves generated near the endwalls do not reach.
Therefore in the basic state the interface could be considered as plane and horizontal.

However, the liquid in the central part of the cavity can't stay at rest, and
the velocities of the liquid motion in the bottom and upper layers must be different.
Indeed, since the densities of liquids are different, then pressures gradients
would be different if the velocities in the layers are equal. The latter is in
contradiction with the normal stress balance condition at the interface.
The only exception might be the case of liquids which stay in rest in laboratory
reference frame, but such a situation is impossible because of the 
incompressibility conditions.
To prove this, let us consider the cross-section of the cavity by the
vertical plane $S_{-+} = S_{-}+S_{+}$, attached to the cavity (Fig. 1).
Then liquid volume $V = V_{r-}+V_{r+}$ to the right (for example) of the 
cross-section is constant. By virtue of the incompressibility the latter means that
in the cavity frame the full liquid consumption through the $S_{-+}$ must be zero:
$$
\forall x: \qquad \int \limits_{-h_{1}}^{\xi} v_{1x} \; dz
+ \int
\limits_{\xi}^{h_{2}} v_{2x} \; dz= 0 \eqno{(2.0)}
$$

\noindent where $\xi$ is the vertical coordinate of the interface.
We stress that the FULL consumption across any vertical cross-section is zero. 
At the same time the consumptions of individual liquids can be non-zero if the flow 
is unsteady because the interface is not at rest near the vertical endwalls,
therefore the volume of each liquid to the right of the cross-section $S_{-+}$
is time-dependent. If the liquids stay in rest in the laboratory frame, then
in (2.0) the velocities $v_{1x}$ and $v_{2x}$ are equal
and (2.0) could not be satisfied. Thus, despite of the statement "in long cavity  
the interface waves generated near the endwalls do not penetrate into
the central part of the cavity", the motion in the central part is always
present and all of the above discussion gives the opportunity for the
approximation of the infinite horizontal layer. Moreover, the condition
(2.0) must be, of course, juncted.

The concrete flow in the inviscid case is obtained in Section 3, 
the flow in viscous case
-- in Section 4.2.
We stress as early as in this Section that the basic flow, as is obvious
from (2.0), is of counter-flow type, that makes us to expect the rise of 
Kelvin--Helmholtz
instability, while the periodic character of the flow allows to expect
the parametric resonance effects.

\begin{figure}[t]
\setlength{\unitlength}{1mm}\thicklines
\begin{center}
\begin{picture}(120,50)
\put(30,53){rigid wall}
\put(30,-5){rigid wall}
\put(20,0){\line(1,0){100}}
\put(122,0){\line(1,0){2}}
\put(126,0){\line(1,0){2}}
\put(130,0){\line(1,0){2}}
\put(132,0){\line(0,1){50}}
\put(20,50){\line(1,0){100}}
\put(122,50){\line(1,0){2}}
\put(126,50){\line(1,0){2}}
\put(130,50){\line(1,0){2}}
\put(20,15){\line(1,0){100}}
\put(122,15){\line(1,0){2}}
\put(126,15){\line(1,0){2}}
\put(130,15){\line(1,0){2}}
\put(18,15){\line(-1,0){2}}
\put(14,15){\line(-1,0){2}}
\put(10,15){\line(-1,0){2}}
\put(18,0){\line(-1,0){2}}
\put(14,0){\line(-1,0){2}}
\put(10,0){\line(-1,0){2}}
\put(18,50){\line(-1,0){2}}
\put(14,50){\line(-1,0){2}}
\put(10,50){\line(-1,0){2}}
\put(8,0){\line(0,1){50}}
\put(94,0){\line(0,1){50}}
\put(87,25){$S_{+}$}
\put(87,5){$S_{-}$}
\put(107,30){$V_{r+}$}
\put(103,6.5){$V_{r-}$}
\put(60,18){interface}
\put(52,2){1}
\put(52,37){2}
\put(20,15){\vector(0,1){20}}
\put(20,15){\vector(1,0){20}}
\put(20,15){\vector(1,1){15}}
\put(16.5,33){z}
\put(38,16.5){x}
\put(35,27){y}
\put(20,15){\vector(0,1){10}}
\put(20,15){\vector(1,0){10}}
\put(16,22){$\vec \gamma$}
\put(28,9){$\vec j$}
\put(19,12.5){o}
\put(134,15){\vector(0,1){35}}
\put(134,50){\vector(0,-1){35}}
\put(136,37){$h_{2}$}
\put(134,0){\vector(0,1){15}}
\put(134,15){\vector(0,-1){15}}
\put(136,7){$h_{1}$}
\put(-10,25){\vector(1,0){15}}
\put(5,25){\vector(-1,0){15}}
\put(-9,27){$a\cos{\omega t}$}
\end{picture}\\
\bigskip
\qquad Fig. 1. 
\end{center}
\end{figure}

In the cavity frame, the Navier--Stokes equations for the
motion could be obtained from the standard equations written in the laboratory
frame by considering the total acceleration, i.e.
$$
-g\vec \gamma \rightarrow -g\vec \gamma + a\omega^{2}\vec j\cos{\omega t}.
$$

\noindent And so, the full Navier--Stokes equations read
$$
{\frac{\partial \vec v_{\beta} }{{\partial t}}} + ( \vec v_{\beta}\nabla
)\vec v_{\beta} = -{\frac{1}{{\rho_{\beta}}}} \nabla p_{\beta} +
\nu_{\beta}\triangle\vec v_{\beta} - g\vec \gamma + a\omega^{2}\vec j\cos {%
\omega t} \eqno{(2.1)}
$$

\bigskip
\noindent and the continuity equations for the two layers read

$$
div\ \vec v_{\beta} = 0 \eqno{(2.2)}
$$

\bigskip
\noindent where the subscript $\beta = 1, 2 $ refers to the lower and upper
layers,
respectively; the other notation being conventional.

No-slip boundary conditions for the velocity are imposed on the rigid walls of the cavity

$$
{\rm at} \ z = -h_{1}: \qquad \vec v_{1} = 0 \eqno{(2.3)}
$$

$$
{\rm at} \ z = h_{2}: \qquad \vec v_{2} = 0 \eqno{(2.4)}
$$

\bigskip

\noindent and on the interface $\; z = \xi(x,y,t) \;$, the following
conditions are satisfied:

\medskip
\noindent - stress balance
$$
(p_{1} - p_{2})n_i = (\sigma^{(1)}_{ik} - \sigma^{(2)}_{ik})n_{k} + \alpha
n_i\;{\it div}\; \vec n \eqno{(2.5)}
$$

\medskip
\noindent - velocity continuity
$$
\vec v_{1} = \vec v_{2} \eqno{(2.6)}
$$

\medskip
\noindent - kinematic condition
$$
{\frac{\partial \xi }{{\partial t}}} + (\vec v_{1} \nabla)\xi = \vec v_{1} \cdot
\vec \gamma \eqno{(2.7)}
$$

\bigskip

\noindent In (2.5 - 2.7),
$\alpha$ is the coefficient of the surface tension,
$\sigma^{(\beta)}_{ik}$ are the viscous stress tensors:
$$
\sigma^{(\beta)}_{ik} = {\frac{\partial v_{\beta , i }}{{\partial x_k}}}
+ {\frac{\partial v_{\beta , k }}{{\partial x_i}}} \ .
$$
The integral condition of zero full consumption reads
$$
\forall x: \qquad \int \limits_{-h_{1}}^{\xi}\vec v_{1} \cdot \vec j \; dz
+ \int
\limits_{\xi}^{h_{2}}\vec v_{2} \cdot \vec j \; dz= 0 \eqno{(2.8)}
$$

\bigskip
\bigskip
\noindent {\large {\bf 3}} \quad {\large {\bf Inviscid approximation}}
\bigskip

After omitting the viscous term
the Navier--Stokes equations (2.1) read:

$$
{\frac{\partial \vec v_{\beta} }{{\partial t}}} + ( \vec v_{\beta}\nabla
)\vec v_{\beta} = -{\frac{1}{{\rho_{\beta}}}} \nabla p_{\beta} - g\vec
\gamma + a\omega^{2}\vec j\cos {\omega t} \eqno{(3.1)}
$$

\medskip
\noindent The continuity equations remain unaltered:
$$
div\ \vec v_{\beta} = 0 \eqno{(3.2)}
$$

\medskip
\noindent The impermeability boundary conditions are imposed
on the rigid walls instead of the no-slip conditions:
$$
{\rm at} \ z = -h_{1}: \qquad \vec v_{1} \cdot \vec \gamma = 0 \eqno{(3.3)}
$$
$$
{\rm at} \ z = h_{2}: \qquad \vec v_{2} \cdot \vec \gamma = 0 \eqno{(3.4)}
$$

\bigskip

\noindent On the interface $\; z = \xi(x,y,t) \;$ the following conditions
are imposed:

\medskip
\indent  - normal stress balance
$$
p_{1} - p_{2} = \alpha \;{\it div}\; \vec n \eqno{(3.5)}
$$

\medskip
\indent  - continuity of normal components of velocity
$$
\vec v_{1}\cdot \vec n = \vec v_{2} \cdot \vec n \eqno{(3.6)}
$$

\medskip
\indent  - kinematic condition
$$
{\frac{\partial \xi }{{\partial t}}} + (\vec v_{1} \nabla)\xi = \vec v_{1} \cdot
\vec \gamma \eqno{(3.7)}
$$

\bigskip
\noindent The zero full consumption condition (2.8) remains unaltered.

\bigskip

The problem (3.1) - (3.7), (2.8) admits a solution of the form:

$$
\vec V_{\beta} = U_{\beta}\vec j \sin{\omega t} \eqno{(3.8)}
$$

\medskip
\noindent where
$$
U_{1} = a\omega{\frac{h_{2}(\rho_{1}-\rho_{2}) }{{h_{1}\rho_{2} +
h_{2}\rho_{1}}}}=a\tilde U_1  \ \ {\rm and} \ \
U_{2} = - a\omega{\frac{h_{1}(\rho_{1}-\rho_{2}) }{{h_{1}\rho_{2} +
h_{2}\rho_{1}}}}=a\tilde U_2 \eqno{(3.9)}
$$

\bigskip
\noindent The corresponding pressure field is given by the expression

$$
p_{\beta} = -\rho_{\beta}gz + a\omega^{2}\rho_{1}\rho_{2}{\frac{h_{1}+h_{2}
}{{h_{1}\rho_{2} + h_{2}\rho_{1}}}} x \cos{\omega t} \eqno{(3.10)}
$$

\bigskip
\noindent The above solution corresponds to a plane-parallel, unsteady,
counter flow which keeps the interface flat ($\;\xi = 0\;$).
Note that in the case of equal densities, $\; \vec V_{1} \;$ and
$\;\vec V_{2} \;$ tend to zero, i.e. fluids stay at rest in the cavity frame.
In the case $\;\rho_{2} \ll \rho_{1}\;$, the lower fluid stays at rest in the
laboratory frame ($\; U_{1} = a\omega\;$).

Thus, in the inviscid case we deal with the stability problem for the
interface between two counter flows, and the difference in the flows velocities
is a periodic function of time.
Linear stability problem could be reduced to the ordinary differential
equation for the amplitude $\; \xi(t) \;$ of the interface displacement from
quasi-equilibrium horizontal position:

$$
(F_{1} + F_{2}){\frac{d^{2}\xi }{{dt^{2}}}} + 2ik{\frac{d\xi }{{dt}}}%
(F_{1}U_{1} + F_{2}U_{2})\sin{\omega t} +
$$
$$
+\xi(\alpha k^{3} + (\rho_{1} - \rho_{2})gk + i(F_{1}U_{1} +
F_{2}U_{2})k\omega\cos{\omega t} - k^{2}(F_{1}U_{1}^{2} +
F_{2}U_{2}^{2})\sin^{2}{\omega t}) = 0 , \eqno{(3.11)}
$$
$$
{\rm with} \qquad F_{1} =\rho_{1}\coth(kh_{1}) \qquad {\rm and} \qquad
\ F_{2} = \rho_{2}\coth(kh_{2})
$$

\noindent In (3.11), $k$ is the wavenumber of the disturbances, which determines
the periodicity of the solution along vibration axis.
It is straightforward to show that stability problem admits the analogue
of Squire's theorem~\cite{squire}. The latter gives the possibility of
considering only plane (2D) disturbances.  

It is convenient to eliminate from (3.11) the term which contains the first
order derivative in $\; \xi \;$. To do so, we make the change of
variable~\cite{shotz}

$$
\xi(t) = Y(t)e^{i\Phi(t)} \eqno{(3.12)}
$$

\medskip
\noindent where

$$
\Phi = {\frac{k }{{\omega}}}{\frac{F_{1}U_{1} + F_{2}U_{2} }{{F_{1} + F_{2}}}%
}\cos{\omega t} \eqno{(3.13)}
$$

\bigskip
\noindent (since $\Phi$ is real, then $\xi$ and $Y$ are equal
via a modulus, i.e. $\xi$ and $Y$ are equivalent from the point of
view of stability). This results in the standard Mathieu equation for $Y$:

$$
{\frac{d^{2}Y }{{dt^{2}}}} + ( A - Q\cos^{2}{t} )Y = 0 \eqno{(3.14)}
$$

\bigskip
\noindent The equation (3.14) is in dimensionless form, with the following
reference quantities for time and length:
$$
t^{*} = 1/\omega \qquad {\rm and} \qquad \ l^{*}= l_c = [\alpha/\left(g(\rho_{1}
- \rho_{2})\right)]^{1/2}
$$

\bigskip
\noindent where $l_c$ is the capillary length, and with the following notation:
$$
Q = {\frac{4B_{v}k^{2} }{{We_1}}}{\frac{\rho \coth(kH_{1}) \coth(kH_{2})
}{{(\rho
\coth(kH_{1})+\coth(kH_{2}))^{2}}}}{\frac{(H_{1}+H_{2})^{2}(\rho-1)^{2} }{{%
(H_{1}+H_{2}\rho)^{2}}}} \eqno{(3.15)}
$$

$$
A = {\frac{k(1+k^{2}) }{{We_1}}}{\frac{\rho-1 }
{{\rho \coth(kH_{1})+\coth(kH_{2})}}%
} \eqno{(3.16)}
$$

\bigskip
\noindent In formulas (3.15) and (3.16), $We_1
= \omega^{2}l_{c}/g$ is Weber number based on the capillary length,
$\rho = \rho_{1}/\rho_{2}$ is densities ratio, $H_{1} = h_{1}/l_c$ and
$H_{2} = h_{2}/l_c$ are dimensionless layer thicknesses.
We also introduced the dimensionless parameter $B_{v}$, characterizing
the vibrations intensity:

$$
B_{v} = {\frac{a^{2}\omega^{2} }{{4}}}\biggl({\frac{\rho_{1} - \rho_{2} }{{%
g\alpha}}}\biggr)^{1/2}.
$$

\bigskip
The solutions of the equation (3.14), which correspond
to the neutral boundary, form two classes: $Y_{+}$, having period $2\pi$
(harmonic disturbances) and $Y_{-}$, having
period $\pi$ (subharmonic disturbances). The solutions of
subharmonic type are antiperiodic, that means that they change the sign
for a shift of $\pi$. 

We plot the boundaries of
the instability regions (obtained by numerical integration of the Mathieu equation (3.14))
in terms of the two parameters, $B_v$ and $k$.
The neutral curves shown in Fig. 2, 3 correspond to Weber numbers equal 
respectively to $We_1=10$ and $We_1=100$ ($H_1=H_2=1, \rho=2$).

The left curve in this Figures bounds Kelvin--Helmholtz instability region.
In the referred case $H_1=H_2=1$ the most dangerous are longwave disturbances with $k=0$.
In~\cite{lyubimov} was investigated the case $We_1 \rightarrow \infty$
with finite heights of the layers. It was found that if $H_1=H_2=H$ then
the transition from the longwave instability to the finite wavelength 
instability occurs at $H=H_*=\sqrt 3$. It is straightforward to analyse 
the case of finite $We_1$. Note, that if $k$ is small,
then $Q$ and $A$ are small also, and could be expanded into power series
in $k$. The result is that at $H < H_*$ the longwave instability
takes place at any values of $We_1$, but at $H > H_*$ only for

$$
We_1 < {3 \over 8}{H \over {H^2-3}}{\rho-1 \over {\rho+1}}
\eqno{(3.17)}
$$

\bigskip
In the instability region discussed, the instability mechanism is, in fact,
the same as in the standard case of Kelvin--Helmholtz instability: the velocity
grows up and the pressure falls down over the elevations of the interface.
The consequence of this is the reducing of the effective stiffness
of the system and, starting from some threshold, the rise of the instability.
As shown in~\cite{lyubimov}, in the case of high frequencies of the vibration 
the nonlinear
developpment of such instability results in the formation of a quasi-stationary 
wave relief on the interface. 

However, periodic changes of the velocity can result
not only in the average effect, but in the resonance amplification of 
eigenoscillations. Inasmuch as in the absence of the dissipation the 
eigenoscillations are not damped, then the instability will take place
at infinitely small amplitude of the vibration (if the synchronism condition
is satisfied).
On Fig. 2, 3 the regions of parametric instability approach the abscissa as narrow ends
(''tongues''); the points $\;k_n\;$ at which the $\;n$-th region of
instability is in contact with
the $\;k$-axis can be evaluated from the equation $A = n^2, n=1,2,...$.
As the 
vibration frequency grows, i.e. as Weber number increases, the points $k_n$ shift
to the shortwave region (as seen from Fig. 2,3) and in the limit 
$We_1 \rightarrow \infty$ they are displaced into infinity.
Thus, at high frequencies of the vibration the parametric instability
can take place only for shortwave disturbances but, for such disturbances,
the viscosity could not be ignored.
It has to be expected that for big values of Weber number only the quasi-static
instability must be observed, with mechanism of excitation which is 
weakly sensitive to the viscosity.

\bigskip
\bigskip
\noindent {\large {\bf 4}} \quad {\large {\bf 2D linear stability problem
for viscous fluids}} \bigskip

To simplify the following presentation, we assume in this Section that the
two layers
are of equal height $\; h_{1} = h_{2} = h. \;$ In addition, we
non-dimensionalize the problem using the scales
$$
t^{*} = 1/\omega; \qquad l^{*} = h; \qquad u^{*} = a\omega; \qquad p^{*} =
\rho_{2}ha\omega^{2}
$$

\bigskip
\bigskip
\noindent {\bf 4.1  Problem formulation}\bigskip

The following dimensionless equations replace the equations (2.1) - (2.8):

$$
{\frac{\partial \vec v_{\beta} }{{\partial t}}} + A( \vec v_{\beta}\nabla
)\vec v_{\beta} = -R_{\beta} \nabla p_{\beta} +
\Omega_{\beta}^{-1}\triangle\vec v_{\beta} - G_o A^{-1}\vec \gamma + \vec
k\cos {t} \eqno{(4.1)}
$$

$$
div\ \vec v_{\beta} = 0 \eqno{(4.2)}
$$

$$
{\rm at} \  z = -1: \qquad \vec v_{1} = 0 \eqno{(4.3)}
$$

$$
{\rm at} \  z = 1: \qquad \vec v_{2} = 0 \eqno{(4.4)}
$$

$$
{\rm at} \ z = \xi(x,y,t): \qquad [p]n_i =
(\Omega_{1}^{-1}\rho\sigma_{ik}^{(1)} -
\Omega_{2}^{-1}\sigma_{ik}^{(2)})n_{k} + A^{-1}We_2^{-1}n_i\; div\ \vec n
\eqno{(4.5)}
$$

$$
[\vec v] = 0 \eqno{(4.6)}
$$

$$
{\frac{1 }{{A}}}{\frac{\partial \xi}{{\partial t}}} + (\vec v_{1}\nabla) \xi
= \vec v_{1} \cdot \vec \gamma \eqno{(4.7)}
$$

$$
\forall x: \qquad \int \limits_{-1}^{\xi}\vec v_{1} \cdot \vec j \; dz + \int
\limits_{\xi}^{1}\vec v_{2} \cdot \vec j \; dz= 0 \eqno{(4.8)}
$$

\bigskip
\noindent
Here and below, the quantity jump across the interface is denoted
by square brackets, for example $[p] = p_{1} - p_{2}$,
and the following parameters are introduced: \\
$A=ah^{-1}$; $\Omega_{\beta}=h^{2}\omega \nu_{\beta}^{-1}$;
$G_o=g/(h\omega^2)$; $We_2=\rho_2 h^3\omega^2/\alpha$;
$\rho = \rho_{1} / \rho_{2}$; $R_{1} = 1/\rho$ and $R_{2} = 1$.

\bigskip
\noindent So, the system of equations governing the fluid motion
contains 7 dimensionless parameters:
$A$ is dimensionless amplitude, $\Omega_{\beta}$ are dimensionless frequencies,
the parameter $G_o$, $We_2$ is a new Weber number,
$k$ is the wave number and $\rho$ is the densities ratio.

\bigskip
\bigskip
\noindent {\bf 4.2  Solution method}\bigskip

The system of equations (4.1) - (4.8) admits a solution of the form

$$
\vec V_{\beta} = U_{\beta}(z)\vec j \exp(it) + C.C. \eqno{(4.9)}
$$

$$
P_{\beta} = -G_o A^{-1}R_{\beta}^{-1}z + x(S\exp(it) + C.C.), \qquad S =
const. \eqno{(4.10)}
$$

\noindent The above solution corresponds to the plane-parallel unsteady counter flow 
which
keeps the interface flat ($\;\xi =0\;$) (to be compared with (3.8)-(3.10) in the
inviscid case).
The function $\;U_\beta (z)\;$ could be found analytically, but its expression
is too complicated, so it was determined numerically.
Figure 4 represents the velocity profiles at four time instants: (a) $t=0$,
(b) $t=\pi /2$, (c) $t=\pi $, (d) $t=3\pi /2$. The direction of the flow
changes every half period of the external forcing. Note the thin
boundary layers at both sides of the interface and near the walls ( Figs.
4a, 4c).

For the reasons discussed in Section 3, we investigate the stability of the
basic state (4.9), (4.10) with respect to periodic (in time and space) $2D$
disturbances. If the vibration frequency is high enough, then thin boundary
layers occur at the interface (on both sides) and near the two horizontal
walls, making the task of numerical solution particularly complicated.

After linearizing near the solution (4.9) and (4.10), the Navier--Stokes
equations for
small disturbances $\vec u_{\beta}$ and $p_{\beta}$ are:

$$
{\frac{\partial \vec u_{\beta} }{{\partial t}}} + A( \vec V_{\beta}\nabla
)\vec u_{\beta} + A( \vec u_{\beta}\nabla )\vec V_{\beta} = -R_{\beta}\nabla
p_{\beta} + \Omega_{\beta}^{-1}\triangle\vec u_{\beta} \eqno{(4.11)}
$$

$$
div\ \vec u_{\beta} = 0 \eqno{(4.12)}
$$

\bigskip
\noindent
Boundary conditions for the system (4.11) - (4.12) read:

$$
{\rm at} \ z = -1: \qquad \vec u_{1} = 0 \eqno{(4.13)}
$$

$$
{\rm at} \ z = 1: \qquad \vec u_{2} = 0 \eqno{(4.14)}
$$

$$
{\rm at} \ z = 0: \qquad
A^{-1}\left((1-\rho)G_o\xi + We_2^{-1} \triangle
\xi\right)n_i +[p]n_i = (\Omega_{1}^{-1}\rho\sigma_{ik}^{(1)} -
\Omega_{2}^{-1}\sigma_{ik}^{(2)})n_{k} \eqno{(4.15)}
$$

$$
[\vec u] = -\left[{\frac{\partial \vec V }{{\partial z}}}\right]\xi
\eqno{(4.16)}
$$

$$
{\frac{1 }{{A}}}{\frac{\partial \xi }{{\partial t}}} + (\vec V_{1}\nabla) \xi
= u_{1z} \eqno{(4.17)}
$$

\bigskip
\noindent
This is valid as long as the deformation $\; \xi \;$ is small compared to
the wavelength of the instability.

It is easy to see that the problem is uniform in $x$-direction. This allows us
to consider only normal disturbances.
The linear stability problem is now summarized as:

$$
{\frac{\partial u_{\beta} }{{\partial t}}} +ikAV_{\beta}u_{\beta} +
w_{\beta}A{\frac{\partial V_{\beta} }{{\partial z}}} = -ikR_{\beta}p_{\beta}
+ \Omega_{\beta}^{-1}( {\frac{\partial^{2} }{{\partial z^{2}}}} - k^{2} )
u_{\beta} \eqno{(4.18)}
$$

$$
{\frac{\partial w_{\beta} }{{\partial t}}} + ikAV_{\beta}w_{\beta} =
-R_{\beta}{\frac{\partial p_{\beta} }{{\partial z}}} + \Omega_{\beta}^{-1}( {%
\ \frac{\partial^{2} }{{\partial z^{2}}}} - k^{2} ) w_{\beta} \eqno{(4.19)}
$$

$$
iku_{\beta} + {\frac{\partial w_{\beta} }{{\partial z}}} = 0 \eqno{(4.20)}
$$

$$
{\rm at} \ z = -1: \qquad u_{1} = w_{1} = 0 \eqno{(4.21)}
$$

$$
{\rm at} \ z = 1: \qquad u_{2} = w_{2} = 0 \eqno{(4.22)}
$$

$$
{\rm at} \ z = 0: \qquad
A^{-1}\xi\left((1-\rho)G_o - We_2^{-1} k^{2}\right) +[p] + 2
\left(\Omega_{2}^{-1}{\frac{\partial w_{2} }{{\partial z}}} -
\Omega_{1}^{-1}\rho{\ \frac{\partial w_{1} }{{\partial z}}}\right) = 0
\eqno{(4.23)}
$$

$$
- {\frac{\Omega_2 }{{\Omega_1 }}}\rho\left(ikw_1+{\frac{\partial u_{1} }{{%
\partial z}}} +\xi{\frac{\partial^{2} V_{1} }{{\partial z^{2}}}}\right)+
ikw_2+ {\frac{\partial u_{2} }{{\partial z}}} +\xi {\frac{\partial^{2} V_{2}
}{{\partial z^{2}}}} = 0 \eqno{(4.24)}
$$

$$
[u] = -\left[{\frac{\partial V }{{\partial z}}}\right]\xi \eqno{(4.25)}
$$

$$
[w] = 0 \eqno{(4.26)}
$$

$$
{\frac{1 }{{A}}}{\frac{\partial \xi }{{\partial t}}} + ikV_{1}\xi = w_{1}
\eqno{(4.27)}
$$

\bigskip

All fields could be expanded in time in Fourier series of the form

$$
f_{\beta}(z,t) = \sum_{n=-\infty}^{\infty}f_{n\beta}(z)\exp(int) + C.C.,
$$

\bigskip
\noindent where $n$ might be taken integer, as well as half-integer.
By keeping only a finite number of terms of the expansion in
both fluids we get $2$-point boundary value problems (BVP's,
coupled through the conditions (4.23) - (4.27)), for the system
of ordinary differential equations
for the complex amplitudes $p_{n\beta}(z)$, $u_{n\beta}(z)$, $w_{n\beta}(z)$
(note that $\xi_{n}(z)$ may be eliminated from (4.23)-(4.27)).
The coefficients in the equations depend on $z$, therefore the system
could be resolved only numerically.  

One important feature
of the resulting ODE's systems is that integer and
half-integer harmonics are not coupled (since the basic flow
(4.9) introduce only a shift of $\pm 1$ for $n$
in (4.18)-(4.27)).
So, finally, the corresponding BVP's can be solved independently.
The solutions of the BVP's
involving integer harmonics have the same period as that of the forcing
vibration (harmonic, or synchronous case) and the solutions of the BVP's
involving half-integer harmonics have a period twice as that of the forcing
vibration (subharmonic, or asynchronous case).

BVP's, in turn, allow the reduction to initial values problems (IVP's).
We focus our attention on the synchronous case, since asynchronous solutions
were not detected in the numerical investigation.
The numerical procedure is described
in~\cite{gershuni}.

\bigskip
\bigskip
\noindent {\large {\bf 5}} \quad {\large {\bf Numerical results}} \bigskip

The linear stability of the flat interface has been
calculated using 21 basis functions $f_{n}(z)$ in both cases
(synchronous and asynchronous).
Ten to twelve basis functions are enough to get sufficiently accurate
numerical results if the dimensionless frequency is not high
($\Omega < \approx 250$). The number of basis functions has to be increased
if $\Omega > 250$.

In Figure 5, we show the neutral stability curves which divide
the $(k,A)$-plane into regions of stable solutions, and regions
(tongues) of unstable (exponentially growing in time) solutions.
The values of the nondimensional parameters are
$\rho$=2, $G_o$=0.16 and $We_2$=6.25.
Only synchronous solutions were detected. This situation is not unique 
(for the review and discussion
see~\cite{or}).
Asymptotic "inviscid" curves are plotted with solid lines, while the curves
obtained numerically for viscous fluids at different values of the
dimensionless frequency $\Omega_1=\Omega_2=\Omega$, are plotted with
dashed lines (note that  $\Omega_1=\Omega_2$ correspond
to $\nu_1 = \nu_2$).

For the high values of $\Omega$ (corresponding to small viscosities)
the numerical curves approach the bottom of the asymptotic tongues,
which correspond to an instability threshold, but lower and lower (Fig. 5).
As $\Omega$ decreases (or the viscosities increase), the instability
threshold
grows (Fig. 6). We state that viscosity has a weak influence on the
Kelvin--Helmholtz instability, but strongly damps the parametric instability
associated with the excitation of capillary-gravity waves. In addition, as
seen in Figures 5,6, at some critical frequency $100 < \Omega_* < 250$, the
boundary of Kelvin--Helmholtz instability region unites with the boundary of
the resonant zone, so that at smaller frequencies there exists a single neutral
curve corresponding to each frequency value. This means that in the case of
rather viscous fluids the difference between the parametric and
non-parametric instabilities vanishes.

\newpage
\noindent {\bf FIGURES CAPTIONS}\\

\begin{tabular} {ll}
Fig. 1.& Problem configuration.\\
& \\
Fig. 2.& Neutral stability diagram (inviscid case), for $We_1=10$.\\
       & Unstable tongues touch the $k$-axis at points $k_n$,\\
       & defined by the equation (3.28). The bigger is the\\
       & tongue's number, the narrower it is.\\
       &  The tongues shift towards shortwave region as the \\
       & vibration frequency increases (next figure).\\
& \\
Fig. 3.& Neutral stability diagram (inviscid case), for $We_1=100$.\\
       & The comparison of this Figure and Figure 2 reveals\\
       & that the influence of changes in the vibration frequency\\ 
       & on Kelvin-Helmholtz instability is very small. \\
& \\
Fig. 4.& Velocity profile at four time instants (viscous case):\\
& (a) $t=0$, (b) $t=\pi /2$, (c) $t=\pi $, (d) $t=3\pi /2$.\\
& Flow direction changes every half period of the \\
& forcing vibration. \\
& \\
Fig. 5.& Neutral stability  $k-A$ diagram (viscous case).\\ 
& Parameters are: $\rho$=2, $G_o$=0.16 and $We_2$=6.25;\\ 
& $\Omega$ is large (250, 360), i.e., viscosity is small .\\
& \\
Fig. 6.& Neutral stability $k-A$ diagram (viscous case). Parameters are\\
& the same as in Fig. 5; $\Omega$ is small (25, 50, 100), i.e.,\\
& viscosity is large. Viscosity makes weak influence on Kelvin--Helmholtz\\
& type instability  but strongly damps the parametric instability.\\
& In the case of rather viscous fluids, the difference between\\ 
& the parametric and non-parametric instabilities vanishes.\\
& \\
\end{tabular}

\end{document}